\documentclass[conference]{IEEEtran}

\usepackage{graphicx,amssymb,amstext,amsmath,epsf,mathtools,indentfirst,bm}
\usepackage{cite}
\usepackage{algpseudocode}
\usepackage{algorithm}
\usepackage{subfigure}

\newtheorem{myprop}{\bf{Proposition}}
\newtheorem{remark}{\bf{Remark}}

\DeclareMathOperator*{\minimize}{\text{minimize}\,}
\DeclareMathOperator*{\st}{\text{subject to}\,}
\DeclareMathOperator*{\maximize}{\text{maximize}\,}

\begin{document}
%
\title{Adaptive Non-myopic Quantizer Design for Target Tracking in Wireless Sensor Networks}

\vspace{-2cm}

\author{
\IEEEauthorblockN{Sijia~Liu\IEEEauthorrefmark{1}, Engin~Masazade\IEEEauthorrefmark{2}, Xiaojing~Shen\IEEEauthorrefmark{3}, Pramod~K.~Varshney\IEEEauthorrefmark{1} \\} 
\IEEEauthorblockA{\IEEEauthorrefmark{1}Syracuse University, NY, 13244, USA, \{sliu17, varshney\}@syr.edu}
\IEEEauthorblockA{\IEEEauthorrefmark{2} Yeditepe University, Istanbul, 34755, Turkey, engin.masazade@yeditepe.edu.tr}
\IEEEauthorblockA{\IEEEauthorrefmark{3} Sichuan University, Chengdu, 610064, China, shenxj@scu.edu.cn}}


\maketitle

\begin{abstract}
In this paper, we investigate the problem of non-myopic (multi-step ahead) quantizer design
for target tracking using a wireless sensor network. 
Adopting the alternative conditional posterior Cram\'er-Rao lower bound (A-CPCRLB) as the optimization metric, we theoretically show that this problem can be temporally decomposed over a certain time window. Based on sequential Monte-Carlo methods for tracking, i.e., particle filters, we design the local quantizer adaptively by solving a particle-based non-linear optimization problem which is well suited for the use of interior-point algorithm and easily embedded in the filtering process. Simulation results are provided to illustrate the effectiveness of our proposed approach.
\end{abstract}


%
\IEEEpeerreviewmaketitle

\section{Introduction}
Wireless sensor networks (WSNs), consisting of a large number of spatially distributed sensors, have been used in a wide range of promising applications such as battlefield surveillance, environment and health monitoring. However, due to the limited communication and energy resources, it is desirable that only quantized data be transmitted from local sensors to the fusion center (FC). 

Quantizer design for target tracking
has been recently studied in the literature \cite{MRV2012,MMP2007,MOHC2011,ORV2008}. In \cite{MRV2012}, the authors employed a static quantizer, first proposed in \cite{NV2006}, to track a moving target, where the optimal quantization thresholds are determined by maximizing the Fisher information about the signal amplitude contained in quantized data. Although this approach is robust and requires minimum prior information about the system, it doesn't yield the optimal solution for tracking scenarios where the target state is random and dynamic. 
An adaptive binary quantizer and uniform quantizer are proposed in \cite{MMP2007} and \cite{MOHC2011}, respectively, where the local quantizers are designed at FC for every time step. In \cite{ORV2008}, a more general framework for designing adaptive identical/non-identical quantizers is presented where the trace of the direct conditional posterior Cram\'er-Rao lower bound (D-CPCRLB \cite{YORP2012}) is minimized. However, D-CPCRLB yields an intractable objective function, which leads to a high computational complexity in optimization. Therefore, to satisfy the real-time operational requirement of the adaptive system, it is essential to seek a tractable objective function and develop an efficient algorithm for quantizer design.

Fisher information matrix (FIM) has been used in \cite{MOHC2011,ORV2008,NV2006,MRV2012} as the performance metric, where the mean square error (MSE) is lower bounded by the inverse of FIM. However, the FIM is generally a matrix. It is important to use a suitable scalar norm of it to obtain a quantity related to the information content. Authors in \cite{ORV2008} employ
the trace of the inverse of FIM. 
The determinant of FIM is used in \cite{MRV2012}, which is inversely proportional to the volume of the uncertainty ellipsoid. In this work, we adopt the trace of FIM as the performance criterion and will theoretically show that maximizing the trace of FIM does not lose the optimality of maximizing FIM in the sense of positive semidefinite cone \cite{Boyd2004_bk,SP_2013}. 
Since the trace operator is linear, we can also show that the trace of FIM yields a tractable objective function for optimization.


For adaptive quantizers proposed in \cite{MMP2007,MOHC2011,ORV2008}, we note that FC is required to feed back the quantization thresholds to local sensors at every time step (a.k.a., the \textit{myopic/greedy} design strategy \cite{ADA2004}). However, continual transmissions might result in data collisions and channel congestion. In order to reduce the amount of communication, we design adaptive quantizers using the \textit{non-myopic} (i.e., multi-steps ahead) strategy, which has drawn recent attention in resource management, e.g., \cite{MNV2012_ciss,MRV2012,ADA2004}.
In general, myopic design has lower computational complexity than the non-myopic case \cite{ADA2004}. However, we will show that the non-myopic quantizer design can be temporally decomposed based on the alternative conditional posterior Cram\'er-Rao lower bound (A-CPCRLB \cite{YORP2012}), which, unlike D-CPCRLB used in \cite{ORV2008}, 
yields a recursive form of the information matrix. With the aid of particle filtering methods\cite{MSN2002}, the problem of non-myopic quantizer design is expressed as a non-linear optimization problem which is easily solved by the interior-point algorithm.



\section{Problem Formulation}
\label{sec: model}
In this paper, the task of the WSN is to monitor a single target moving in a two-dimensional Cartesian coordinate plane. At sampling time $t$, the target state is defined by a $4 \times 1$ dimensional vector $\mathbf x_t = [x_t, y_t, \dot{x}_t, \dot{y}_t]$ where $(x_t,y_t)$ and $(\dot x_t, \dot y_t)$ denote the target location and velocity in the 2D plane, respectively. The target state evolves according to
\begin{align}
\resizebox{.26\hsize}{!}{$
\displaystyle { \mathbf x}_{t+1}= {F}{\mathbf x}_{t} + {\mathbf w}_{t},
$}
\label{eq: state0}
\end{align} 
where $\mathbf w_t \sim \mathcal N(\mathbf 0, Q)$, the state transition matrix $F$ and the process noise covariance $Q$ are given by \cite{MRV2012}
\begin{equation}
\label{eq:Necessary_matrices}
\resizebox{.85\hsize}{!}{
$\displaystyle {F} = \left[
  \begin{array}{cccc}
    1 & 0 & \Delta & 0 \\
    0 & 1 & 0 & \Delta \\
    0 & 0 & 1 & 0 \\
    0 & 0 & 0 & 1 \\
  \end{array}
\right] ~ {Q} = q \left[
  \begin{array}{cccc}
    \frac{\Delta^3}{3} & 0 & \frac{\Delta^2}{2} & 0 \\
    0 & \frac{\Delta^3}{3} & 0 & \frac{\Delta^2}{2} \\
    \frac{\Delta^2}{2} & 0 & \Delta & 0 \\
    0 & \frac{\Delta^2}{2} & 0 & \Delta \\
  \end{array}
\right].$
}
\end{equation}
In (\ref{eq:Necessary_matrices}), $\Delta$ and $q$ denote the sampling interval between adjacent sensor measurements and the process noise parameter, respectively.

We further consider $N$ sensors deployed in a region of interest (ROI) and each of them reports a noisy measurement in the form of signal power \cite{MRV2012,ORV2008}
\begin{equation}
\resizebox{.65\hsize}{!}{$
\displaystyle
\begin{array}{cc}
{y}_t^i = h_t^i ( {\mathbf x}_t )  + { v}_t^i, & 
 h_t^i ( {\mathbf x}_t ) = \sqrt{\frac{P_0}{1+ (d_t^i)^{2}}}
 \end{array}
$}
\label{eq: meas0}
\end{equation}
for $i = 1, 2, \cdots, N$, where $v_t^i \sim \mathcal N (0, \sigma_v^2)$, 
$P_0$ denotes the signal power of the source, 
$d_t^i$ is the distance between the target and the $i$th sensor, $d_t^i = \sqrt{(\mathsf x_i -  x_{t})^2 + (\mathsf y_i - y_{t})^2}$, where $(\mathsf x_i, \mathsf y_i)$ is the position of the $i$th sensor in the 2D plane. 

Each sensor quantizes its measurement to $M$ bits as given below
\begin{align}
\resizebox{.63\hsize}{!}{$
\displaystyle 
u_t^i = \left\{
  \begin{array}{c c}
    0 & \quad -\infty < y_t^i \leq \gamma_{t,1}^i  \\
    \vdots & \quad \vdots \\
    L-1 & \quad \gamma_{t,L-1}^1 < y_t^i < +\infty \\
  \end{array} \right. ,
  $}
  \label{eq: thresh_q}
\end{align}
where $u_t^i$ denotes the quantized measurement of the $i$th sensor at time step $t$, $L = 2^M$, and the vector $\bm \gamma_t^i : = [\gamma_{t,1}^i, \cdots, \gamma_{t,L-1}^i]^T$ corresponds to the quantization strategy of sensor $i$. For notational consistency, let $\gamma_{t,0}^i = -\infty$ and $\gamma_{t,L}^i = \infty$. It is clear from (\ref{eq: thresh_q}) that the probability of a particular quantization output $l$ is 
\begin{align}
\resizebox{.8\hsize}{!}{$
\displaystyle p(u_t^i = l | \mathbf x_t) = Q(\frac{\gamma_{t,l}^i-h_i^i (\mathbf x_t)}{\sigma_v}) - Q(\frac{\gamma_{t,l+1}^i-h_t^i(\mathbf x_t)}{\sigma_v})$},
\label{eq: prob_threshold}
\end{align}
where $Q(\cdot)$ is the complementary distribution function of the standard Gaussian distribution. 
Under the assumption of conditionally independent observations at local sensors, 
the observation likelihood function at time $t$ can be written as 
\begin{align}
\resizebox{.36\hsize}{!}{$
\displaystyle p(\mathbf u_{t}| \mathbf x_{t}) = \prod_{i=1}^N p( u_{t}^i | \mathbf x_{t}),
$}
\label{eq: likelihood_M}
\end{align}
where $\mathbf u_t = [u_t^1, u_t^2, \cdots, u_t^N]^T$ denotes the collection of quantized measurements from $N$ sensors. 

\subsection{Alternative conditional posterior Cram\'er-Rao lower bound (A-CPCRLB)}

A conditional posterior Cra\'mer-Rao lower bound (C-PCRLB) is proposed in \cite{LRV2011} by incorporating the history of actual sensor observations, which can provide a tighter error bound than the conventional PCRLB. Nevertheless, obtaining C-PCRLB is not computationally efficient due to the presence of the auxiliary Fisher information matrix \cite[Thm.\,$1$]{LRV2011}. 
Therefore, Zheng \textit{et al.} in \cite{YORP2012} presented an alternative conditional PCRLB (A-CPCRLB), which is direct and more compact. In this work, we adopt A-CPCRLB as the performance criterion for quantizer design.

Let $\mathbf x_{0:t}$ and $\mathbf u_{1:t}$ denote the state vector and measurements up to time $t$. Then the conditional mean squared error of the state vector $\mathbf x_{0:t}$ is lower bounded by the inverse of the conditional Fisher information matrix (C-FIM) as in \cite{LRV2011}
\begin{align*}
\resizebox{.9\hsize}{!} {$E \{ [\hat {\mathbf x}_{0:t+1} - {\mathbf x}_{0:t+1}] [\hat {\mathbf x}_{0:t+1} - {\mathbf x}_{0:t+1}]^T | \mathbf u_{1:t} \} \geq  J^{-1}(\mathbf x_{0:t+1}|\mathbf u_{1:t})$}.
\end{align*}
Let $ J(\mathbf x_{t+1}|\mathbf u_{1:t})$ be the matrix whose inverse equals the lower-right corner submatrix of $J^{-1}(\mathbf x_{0:t+1}|\mathbf u_{1:t})$. Then the matrix $ J(\mathbf x_{t+1}|\mathbf u_{1:t})$ provides a lower bound on the mean square error (MSE) of estimating $\mathbf x_{t+1}$. As shown in \cite[Corollary\,$1$]{YORP2012}, for the linear Gaussian model (\ref{eq: state0}), the 
C-FIM $ J(\mathbf x_{t+1}|\mathbf u_{1:t})$ can be computed as follows,
\begin{align}
\resizebox{.9\hsize}{!}{$
J_{t+1} \approx \left( Q + F J_{t}^{-1}F^T \right)^{-1} + E_{p_{t+1}^c} \left \{  - \nabla_{\mathbf x_{t+1}}^{\mathbf x_{t+1}}   \mathrm{ln} p (\mathbf u_{t+1}| \mathbf{x}_{t+1})\right \}
$},
\label{eq: A_CBIM}
\end{align}
where for notional simplicity we use $J_{t+1}$ instead of $ J(\mathbf x_{t+1}|\mathbf u_{1:t})$, $\nabla_{\mathbf x}^{\mathbf x}$ is the second-order partial derivative with respect to $\mathbf x$ and $p_{t+1}^c \triangleq p(\mathbf x_{t+1}, \mathbf u_{t+1} | \mathbf u_{1:t})$. Note that the first term in (\ref{eq: A_CBIM}) is the prediction of $J_{t}$ using the state evolution model and the second term 
indicates the information based on the updated measurements $\mathbf u_{t+1}$ at time $t+1$.


\subsection{Non-myopic quantizer design}
For the non-myopic quantizer design, we seek optimal quantizers defined in (\ref{eq: thresh_q}) over the next $T_w$ time steps, $t+1 : t+T_w$, at time instant $t$.
It is clear from (\ref{eq: likelihood_M}) and (\ref{eq: A_CBIM}) that the value of C-FIM $J_{t+\eta}$ relies on quantization thresholds of local sensors
at time $t+\eta$ (denoted by $\bm \gamma_{t+\eta} \triangleq [\bm \gamma_{t+\eta}^1, \cdots, \bm \gamma_{t+\eta}^N]$), the previous C-FIM $J_{t+\eta-1}$, and the conditional distribution $p_{t+\eta}^c$, i.e., $p(\mathbf x_{t+\eta}, \mathbf u_{t+\eta} | \mathbf u_{1:t+\eta-1} )$
, where $\eta \in \{1,2,\cdots, T_w\}$ and $J_t$ is used as
prior information. However, during the design window $t+1:t+T_w$, the conditional PDF $p_{t+\eta}^c$ cannot be obtained exactly for $\eta>1$ since the quantized measurements $\mathbf u_{t:t+\eta-1}$ are not available at time $t$. Therefore, as in \cite{MNV2012_ciss}, the conditional PDF $p_{t+\eta}^c$ is approximated by its prediction $p(\mathbf x_{t+\eta}, \mathbf u_{t+\eta} | \mathbf u_{1:t} )$, which is easily obtained using a particle filter (see more details in Sec. \ref{sec: MC}). 

To determine optimal thresholds $\{ \bm \gamma_{t+\eta}\}_{\eta=1,\cdots, T_w}$
for the next $T_w$ time steps, we pose the optimization problem as given below where we maximize C-FIM at time $t+T_w$,
\begin{equation}
\resizebox{.7\hsize}{!}
{$
\displaystyle
\begin{array}{cl}
\displaystyle \maximize_{\{\bm \gamma_{t+\eta}\}} &  J_{t+T_w} (\bm \gamma_{t+1},\cdots, \bm \gamma_{t+T_w})  \\
\text{subject to} & \gamma_{t+\eta,1}^i < \cdots < \gamma_{t+\eta,L-1}^i \\
& \text{$\eta=1,\cdots, T_w$ and $i = 1, \cdots, N$}
\end{array}
$}
\label{eq: non_mop_CFIM}
\end{equation}
where for notational simplicity we use $\{\cdot\}$ instead of $\{\cdot\}_{\eta = 1,\cdots,T_w}$, $\bm \gamma_{t+\eta}$ is a $(L-1) \times N$ quantizer threshold matrix whose element $\gamma_{t+\eta,l}^i$ represents the $l$th threshold of sensor $i$ at time $t+\eta$, $L$ indicates the number of quantization levels, $N$ is the number of sensors and $T_w$ is the length of time window. 

Note that problem (\ref{eq: non_mop_CFIM}) is a matrix optimization problem which is defined in the positive semidefinite cone \cite{Boyd2004_bk}. Namely, 
if $ \{\bm \gamma_{t+\eta}^* \}$ is an optimal solution, then for an arbitrary feasible solution 
$ \{\bm \gamma_{t+\eta} \}$, $J_{t+Tw}( \{\bm \gamma_{t+\eta}^* \} ) \succeq J_{t+Tw}( \{\bm \gamma_{t+\eta} \})$\footnote{A positive/negative semidefinite matrix $A$ is denoted by $A \succeq  0$ or $A \preceq 0$.}, i.e., the matrix $J_{t+Tw}( \{\bm \gamma_{t+\eta}^* \} ) - J_{t+Tw}( \{\bm \gamma_{t+\eta} \} )$ is positive semidefinite.


Furthermore, the following Proposition shows that the problem (\ref{eq: non_mop_CFIM}) can be equivalently transformed to $T_w$ sub-problems. Each of the subproblems has a scalar objective function in terms of the trace of the Fisher information matrix with respect to the updated measurements.

\begin{myprop}
\label{prop_decompose}
If problem (\ref{eq: non_mop_CFIM}) has an optimal solution, then the solution of (\ref{eq: non_mop_CFIM}) can be equivalently transformed to the solution of
$T_w$ subproblems, i.e.,
\begin{equation}
\resizebox{.8\hsize}{!}
{$
\displaystyle
\begin{array}{cl}
\displaystyle \maximize_{\bm \gamma_{t+\eta}} &  \mathrm{tr}\left( E_{p_{t+\eta}^c} \left \{  - \nabla_{\mathbf x_{t+\eta}}^{\mathbf x_{t+\eta}}   \mathrm{ln} p (\mathbf u_{t+\eta}| \mathbf{x}_{t+\eta})\right \} \right) \\
\text{subject to} &  \gamma_{t+\eta,1}^i < \cdots < \gamma_{t+\eta,L-1}^i, ~ \text{$i = 1, \cdots, N$}
\end{array}
$}
\label{eq: decompose_nmop}
\end{equation}
for $\eta=1,2,\cdots,T_w$, where $\mathrm{tr}(\cdot)$ denotes the trace operator and $p_{t+\eta}^c \approx p(\mathbf x_{t+\eta}, \mathbf u_{t+\eta} | \mathbf u_{1:t} )$.
\end{myprop}

\textbf{Proof:}
See appendix.
\hfill$\blacksquare$



\section{Particle-based non-myopic quantizer design}
\label{sec: MC}


In this section, we will show that the problem (\ref{eq: decompose_nmop}) can be further expressed in a closed form and solved efficiently with the aid of a particle filtering method.



At time step $t+\eta$, by substituting (\ref{eq: likelihood_M}) into (\ref{eq: decompose_nmop}),
the objective function in (\ref{eq: decompose_nmop}) can be written as
\begin{align*}
\resizebox{.8\hsize}{!}{$
\displaystyle
 \sum_{i=1}^N \mathrm{tr} \left( E_{p(\mathbf x_{t+\eta}, u_{t+\eta}^i | \mathbf u_{1:t} )} \left \{ - \nabla_{\mathbf x_{t+\eta}}^{\mathbf x_{t+\eta}}   \mathrm{ln} p ( u_{t+\eta}^i| \mathbf{x}_{t+\eta}) \right \} \right),
$}
\end{align*}
which indicates that seeking the optimal quantizers of $N$ sensors at time $t+\eta$ can be obtained by equivalently solving a sequence of sub-problems, i.e., 
\begin{equation}
\resizebox{.9\hsize}{!}{$
\displaystyle
\begin{array}{cl}
\displaystyle \maximize_{\bm \gamma_{t+\eta}^i} &  \psi(\bm \gamma_{t+\eta}^i) \triangleq \mathrm{tr}\left( E \left \{ - \nabla_{\mathbf x_{t+\eta}}^{\mathbf x_{t+\eta}}   \mathrm{ln} p ( u_{t+\eta}^i| \mathbf{x}_{t+\eta}) \right \} \right) \\
\text{subject to} &  \gamma_{t+\eta,1}^i < \cdots < \gamma_{t+\eta,L-1}^i, 
\end{array}
$}
\label{eq: subsubprob_nonmop}
\end{equation}
for $i=1,2,\cdots,N$.

Using the fact that $u_{t+\eta}^i$, $\mathbf x_{t+\eta}$ and $\mathbf u_{1:t}$ form a Markov chain and the identity for standard Fisher information matrix \cite{HK_bk}
\begin{align}
\resizebox{.75\hsize}{!}{$
\displaystyle
E \left[ \frac{\partial \text{ln}p(u_{t}^i |\mathbf x_{t})}{\partial \mathbf x_{t,r}}  \frac{\partial \text{ln}p(u_{t}^i |\mathbf x_{t})}{\partial \mathbf x_{t,j}} \right] = -E \left[  \frac{\partial^2 \text{ln} p(u_{t}^i |\mathbf x_{t})}{\partial \mathbf x_{t,r} \partial \mathbf x_{t,j}} \right]
$}
\end{align}
where $\mathbf x_{t,r}$ and $\mathbf x_{t,j}$ denote the $r$th entry and $j$th entry of state vector $\mathbf x_t$, the objective function $\psi(\bm \gamma_{t+\eta}^i)$ in (\ref{eq: subsubprob_nonmop}) can be written as
\begin{equation}
\label{eq: obj_subsubprob}
\resizebox{.95\hsize}{!}{$
\displaystyle
\psi(\bm \gamma_{t+\eta}^i)   = \displaystyle \sum_{r = 1}^4  E_{p(\mathbf x_{t+\eta} | \mathbf u_{1:t})} 
\left \{ E_{p( u_{t+\eta}^i|\mathbf x_{t+\eta})} \left[ \left( \frac{\partial \mathrm{ln}p(u_{t+\eta}^i |\mathbf x_{t+\eta})}{\partial x_{t+\eta,r}} \right)^2 \right ] \right \}  
$}
\end{equation}
where the likelihood $p(u_{t+\eta}^i |\mathbf x_{t+\eta})$ is given by (\ref{eq: prob_threshold}). 

We employ a particle based method to compute the posterior PDF. In a SIR filter \cite{MSN2002}, the posterior PDF $p(\mathbf x_t | \mathbf u_{1:t})$ is approximated by a set of particles $\{ \mathbf x_t^s; s = 1,\ldots, N_s\}$ with equal weights $1/N_s$ after the re-sampling process, where $N_s$ is the total number of particles. Thus, the predicted PDF $p(\mathbf x_{t+\eta}|\mathbf u_{1:t})$ in (\ref{eq: obj_subsubprob}) can be obtained by propagating particles $\mathbf x_t^s$ after $\eta$ steps using the state model (\ref{eq: state0}). Then, 
\begin{align}
\resizebox{.54\hsize}{!}{$
\displaystyle
p(\mathbf x_{t+\eta}| \mathbf u_{1:t}) \approx \frac{1}{N_s} \sum_{s= 1}^{N_s}\delta (\mathbf x_{t+\eta} - \mathbf x_{t+\eta}^s).
$}
\label{eq: predic_pdf}
\end{align}

Substituting (\ref{eq: prob_threshold}) 
and (\ref{eq: predic_pdf}) into (\ref{eq: obj_subsubprob}), the optimization problem (\ref{eq: subsubprob_nonmop}) can be written as
\begin{equation}
\resizebox{.7\hsize}{!}{$
\displaystyle
\begin{array} {cl}
\displaystyle \maximize_{\bm \gamma_{t+\eta}^i 
} & 
\psi (\bm \gamma_{t+\eta}^i) = \displaystyle  \sum_{l=0}^{L-1} f(\gamma_{t+\eta,l}^i, \gamma_{t+\eta,l+1}^i) \\
\st &  \gamma_{t+\eta,1}^i < \gamma_{t+\eta,2}^i< \cdots < \gamma_{t+\eta,L-1}^i
\end{array},
$}
\label{eq: subsub_nmop_particle}
\end{equation}
where 
\begin{align*}
\resizebox{.9\hsize}{!}{$
\displaystyle f(\gamma_{t+\eta,l}^i, \gamma_{t+\eta,l+1}^i) = \frac{1}{N_s \sigma_v^2}\sum_{s = 1}^{N_s} \sum_{r = 1}^4 g(\gamma_{t+\eta,l}^i,\gamma_{t+\eta,l+1}^i, \mathbf x_{t+\eta}^s, r),
$}
\end{align*}
and
\begin{equation}
\begin{array}{ll}
& \resizebox{.37\hsize}{!}{$ \displaystyle g(\gamma_{t+\eta,l}^i,\gamma_{t+\eta,l+1}^i, \mathbf x_{t+\eta}^{s}, r) $} \\
= &
\resizebox{.8\hsize}{!}{$
\displaystyle
\frac{ (\frac{\partial h_{t+\eta}^i ( \mathbf x_{t+\eta}^s)}{\partial \mathbf x_{t+\eta,r}})^2 \left[  q ( \frac{\gamma_{t+\eta,l}^i- b_{t+\eta}^{i,s} }{\sigma_v} ) -   q ( \frac{\gamma_{t+\eta,l+1}^i- b_{t+\eta}^{i,s}}{\sigma_v}) \right]^2}{Q(\frac{\gamma_{t+\eta,l}^i - b_{t+\eta}^{i,s}}{\sigma_v}) -Q(\frac{\gamma_{t+\eta,l+1}^i - b_{t+\eta}^{i,s}}{\sigma_v})}
$}
\end{array}
\label{eq: Hfunc}
\end{equation}
with $b_{t+\eta}^{i,s} \triangleq h_{t+\eta}^i (\mathbf x_{t+\eta}^s)$.

It is clear from (\ref{eq: Hfunc}) that 
the objective function of (\ref{eq: subsub_nmop_particle}) is non-linear but differentiable. Therefore, the interior-point algorithm \cite{Boyd2004_bk} is a well-suited optimization tool for solving problem (\ref{eq: subsub_nmop_particle}). The procedure for non-myopic quantizer design under the SIR filtering framework is summarized in Algorithm\,$1$. 
\begin{remark}
It can be seen from (\ref{eq: subsub_nmop_particle}) that the complexity of the non-myopic quantizer design depends on the number of Monte-Carlo particles (i.e., $N_s$). Fewer number of particles would reduce the computation cost but result in worse estimation performance due to the low accuracy of approximating the predicted PDF in (\ref{eq: predic_pdf}). 
Therefore, it is important to investigate the tradeoff between the number of particles and the estimation performance.
\end{remark}

\begin{algorithm}
\caption{\small Adaptive non-myopic quantizer design }
\begin{algorithmic}[1]
\State \small At time $t$, begin with the updated particles $\mathbf x_t^s$ and weights $w_t^s = N_s^{-1}$
\For{$\eta=1,\ldots, T_w$}
\State Propagate particles by $\mathbf x_{t+1}^s = F \mathbf x_t^s + w_t$
\State $p(x_{n+1}|u_{1:n_{-}}) = \frac{1}{N_s}\sum_{s=1}^{N_s} \delta(x_{n+1}-x_{n+1}^s)$
\State Obtain optimal thresholds $\gamma_{t+1}^i$ for $N$ sensors by solving \verb"   "\!\! 
(\ref{eq: subsub_nmop_particle}) for $i=1,\cdots,N$.
\EndFor
\State Feed $\{\gamma_{t+\eta}\}_{\eta=1,\ldots,T_w}$ back to local sensors and update particles by using the corresponding quantized measurement at $t+1, \ldots, t+T_w$.
\end{algorithmic}
\end{algorithm}

\subsection{Binary Quantizers}
For a binary quantizer, i.e., $L=2$, the optimization problem (\ref{eq: subsub_nmop_particle}) becomes \textit{unconstrained}, i.e.,
\begin{equation}
\resizebox{.69\hsize}{!}{$
\displaystyle
\begin{array}{cl}
\displaystyle \maximize_{\gamma_{t+\eta,1}^i} & f(-\infty, \gamma_{t+\eta,1}^i) + f(\gamma_{t+\eta,1}^i, \infty)
\end{array},
$}
\label{eq: subsub_BQ}
\end{equation}
whose optimality condition is presented by the following proposition.

\begin{myprop}
The optimality condition for the binary quantizer of sensor $i$ at time $t+\eta$ can be expressed by a nonlinear equation
\begin{align*}
\resizebox{1.05\hsize}{!}{$
\begin{array}{l}
\displaystyle \sum_{r=1}^4 \sum_{s = 1}^{N_s} \left [ \frac{\partial h_{t+\eta}^i (\mathbf x_{t+\eta}^s)}{\partial \mathbf x_{t+\eta,r}}\right ]^2  \frac{2 (\gamma_{t+\eta,1}^i- b_{t+\eta}^{i,s}) q^2(\frac{\gamma_{t+\eta,1}^i- b_{t+\eta}^{i,s}}{\sigma_v})}{Q(\frac{\gamma_{t+\eta,1}^i - b_{t+\eta}^{i,s}}{\sigma_v})\left[ 1-Q(\frac{\gamma_{t+\eta,1}^i - b_{t+\eta}^{i,s}}{\sigma_v} ) \right ]} \\
\displaystyle +  \sigma_v \sum_{r=1}^4 \sum_{s = 1}^{N_s} \left[ \frac{\partial h_{t+\eta}^i (\mathbf x_{t+\eta}^s)}{\partial \mathbf x_{t+\eta,r}}\right]^2
\frac{q^3(\frac{\gamma_{t+\eta,1}^i- b_{t+\eta}^{i,s}}{\sigma_v}) (2Q(\frac{\gamma_{t+\eta,1}^i - b_{t+\eta}^{i,s}}{\sigma_v})-1)}{Q^2(\frac{\gamma_{t+\eta,1}^i - b_{t+\eta}^{i,s}}{\sigma_v})\left [ 1-Q(\frac{\gamma_{t+\eta,1}^i - b_{t+\eta}^{i,s}}{\sigma_v})\right ]^2} \\
 = 0,
\end{array}
$}
\end{align*}
where $\gamma_{t+\eta,1}^i$ represents the quantization threshold of the $i$th sensor at time $t+\eta$, $b_{t+\eta}^{i,s} \triangleq h_{t+\eta}^i (\mathbf x_{t+\eta}^s)$, and
$h_{t+\eta}^i (\cdot)$ is the measurement model.

\textbf{Proof:} The result can be easily obtained by taking the first-order derivative of (\ref{eq: subsub_BQ}).
$\blacksquare$ \hfil 
\end{myprop} 


\subsection{Identical Quantizers}
It is clear from 
(\ref{eq: subsub_nmop_particle}) that local quantizer design relies on the sensor location and the predicted measurement $h_{t+\eta}^i(\mathbf x_{t+\eta}^s)$, which theoretically verifies the statement in \cite{ORV2008} that the use of identical quantizers at all the sensors leads to performance degradation. On the other hand,  
if we consider a linear measurement model $y_t^i = \mathbf h_t^i\mathbf x_t + v_t^i$ for sensor $i$ at time $t$,
where $\mathbf h_t^1 = \cdots = \mathbf h_t^N$ (e.g., the mean estimation problem in \cite{ABP2013}), it can be shown that the problem of quantizer design
yields identical optimal thresholds for $N$ sensors at every time instant due to the effect of identical sensor observation model on (\ref{eq: prob_threshold}) and (\ref{eq: obj_subsubprob}).

\section{Simulation Results}
\label{sec: Sim}
In our simulations, we consider that $N=9$ sensors are grid deployed
in a $20 \times 20~m^2$ surveillance area.
For target motion in (\ref{eq: state0}), we select the sampling interval $\Delta = 0.5$ seconds and process noise parameter $q \in \{0.1, 2.5 \times 10^{-3}\}$, where the magnitude of $q$ indicates the relative uncertainty regarding the target trajectory (see \cite[Fig\,$3$]{MRV2012} for an example). The initial state distribution of the target is assumed to be Gaussian with mean $\mu_0 = [-8.8,-8.8,1.8,1.8]$ and covariance $\Sigma_0 = \mathrm{diag}[\sigma_0^2, \sigma_0^2, 0.01,0.01]$ where $3 \sigma_0 = 2$. We perform target tracking over $10$ seconds, i.e., $20$ time steps ($T_w \leq 20$).
Sensor measurements are obtained from (\ref{eq: meas0}), where $P_0 = 1000$ and sensor observation noise with $\sigma_v = 0.1$.
We assume that the fusion center has perfect information about the target dynamical model and the noise statistics. 
Observing from simulation results, which are omitted here for brevity, using $1000$ and $50$ particles in target estimation and quantizer design, respectively, provides a suitable tradeoff between the number of particles and tracking performance, which is evaluated in terms of mean square error (MSE) over $100$ trials. 

\begin{figure}[hbt]
\centering
\subfigure[]{
\includegraphics[width=.38\textwidth,height=!]{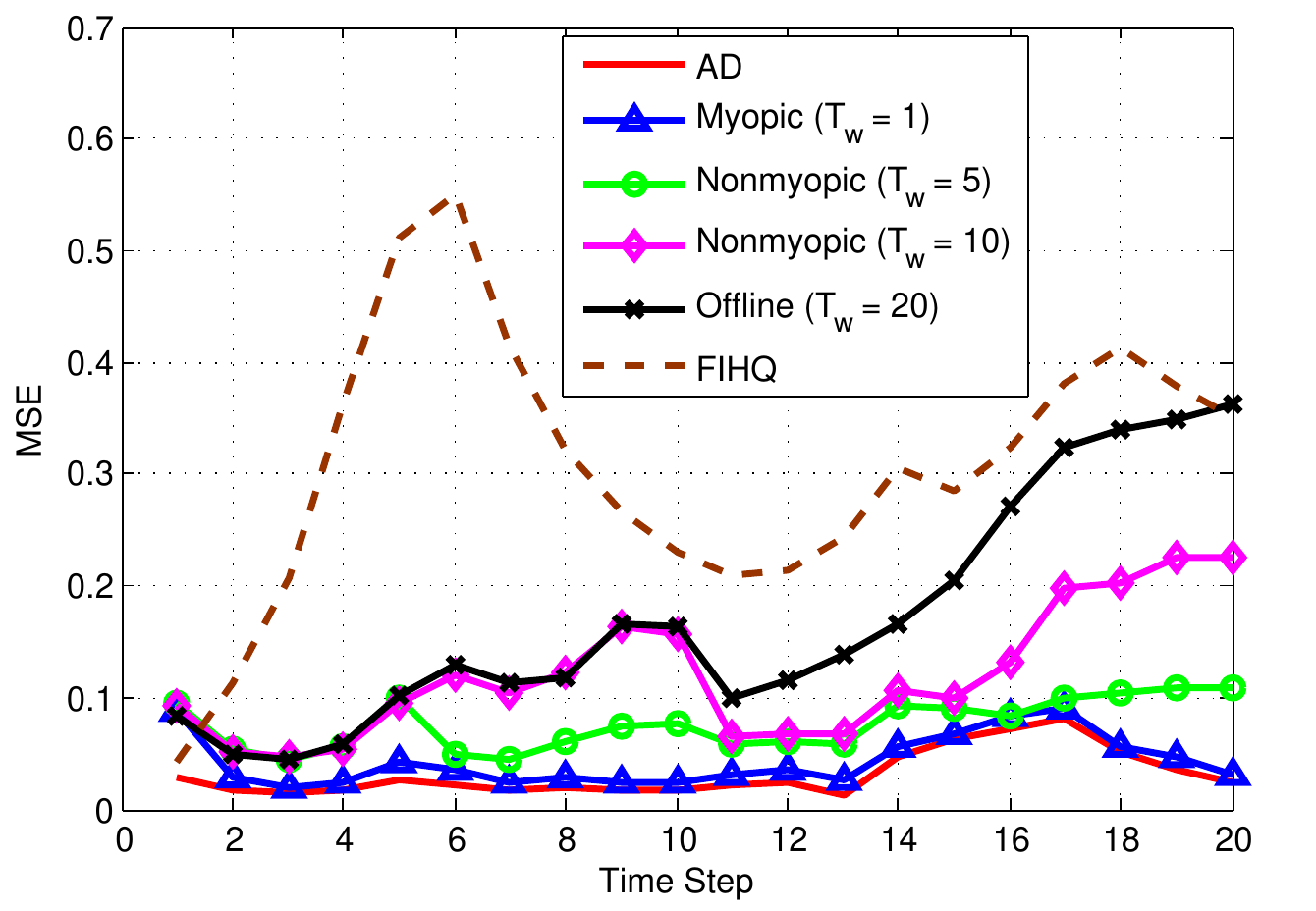} }
\subfigure[]{
\includegraphics[width=.38\textwidth,height=!]{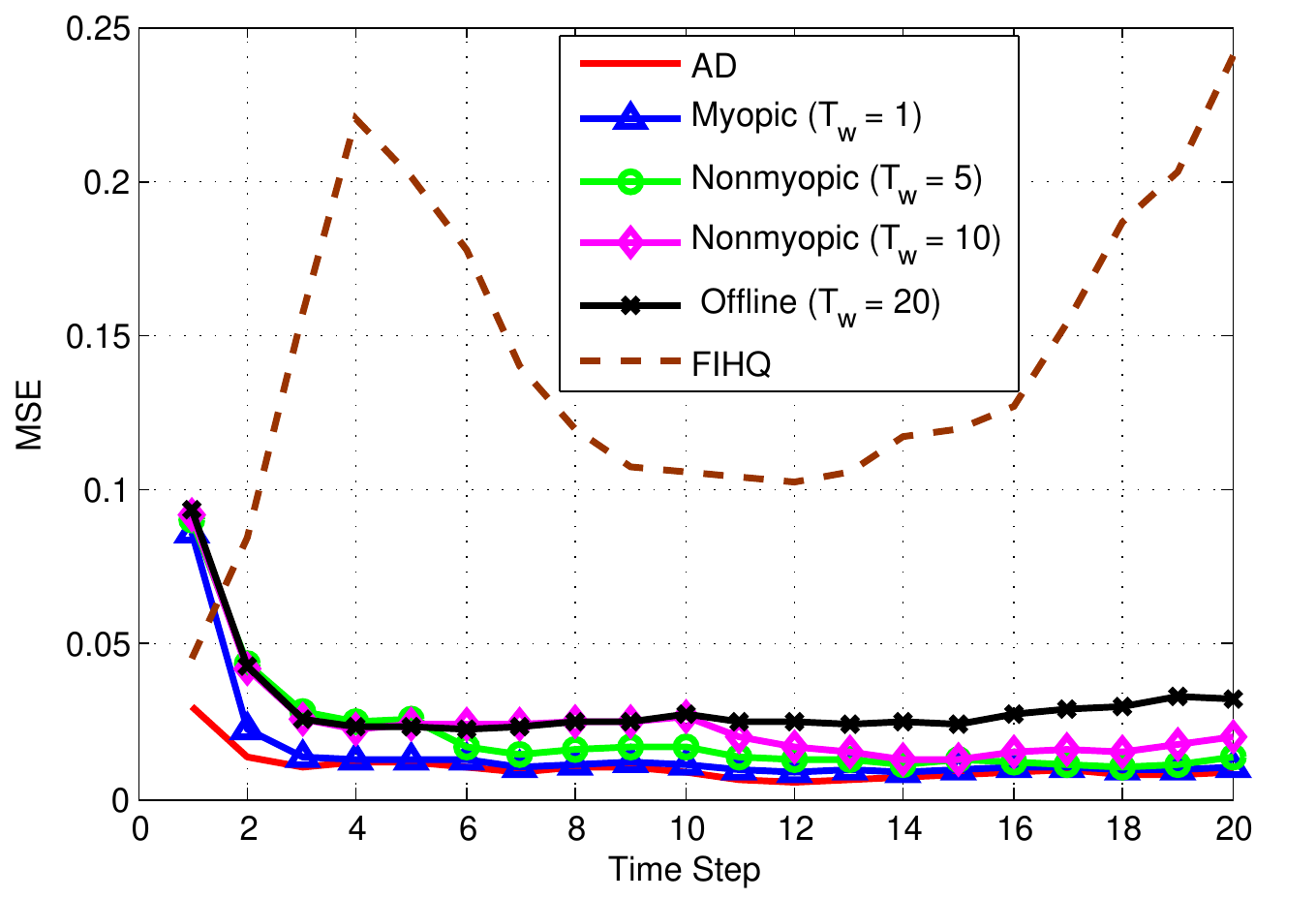} }
\caption{\footnotesize {Tracking performance of the non-myopic quantizer with different sizes of time window (a) $q=0.1$ (b) $q = 2.5 \times 10^{-3}$}}
\label{fig: nonmop}
\end{figure}

In Fig.\,\ref{fig: nonmop}, we demonstrate the tracking performance of the $2$-bits non-myopic quantizer for different time window sizes, i.e., $T_w \in \{1,5,10,20\}$, where the non-myopic design with $T_w = 1$ is equivalent to the myopic design, and the non-myopic quantizer with $T_w = 20$ becomes an offline quantizer since the corresponding A-CPCRLB is calculated offline. For comparison, we also present the tracking performance when using analog data (AD) and quantized data based on the offline Fisher information heuristic quantizer (FIHQ) \cite{NV2006}.  As we can see, our proposed quantization strategy yields better performance than FIHQ. Specifically,
Fig.\,\ref{fig: nonmop}-(a) shows that the estimation performance improves as $T_w$ decreases. This is because for $q = 0.1$, the target trajectory has relatively large uncertainty so that the accuracy of estimation benefits from quantizer design using more sensor measurements. However, Fig.\,\ref{fig: nonmop}-(b) shows that the MSE for all values of $T_w$ lies close to each other since the target trajectory is almost deterministic (and thus predictable) as $q = 2.5 \times 10^{-3}$. 

\begin{figure}[hbt]
\centering
\includegraphics[width=.39\textwidth,height=!]{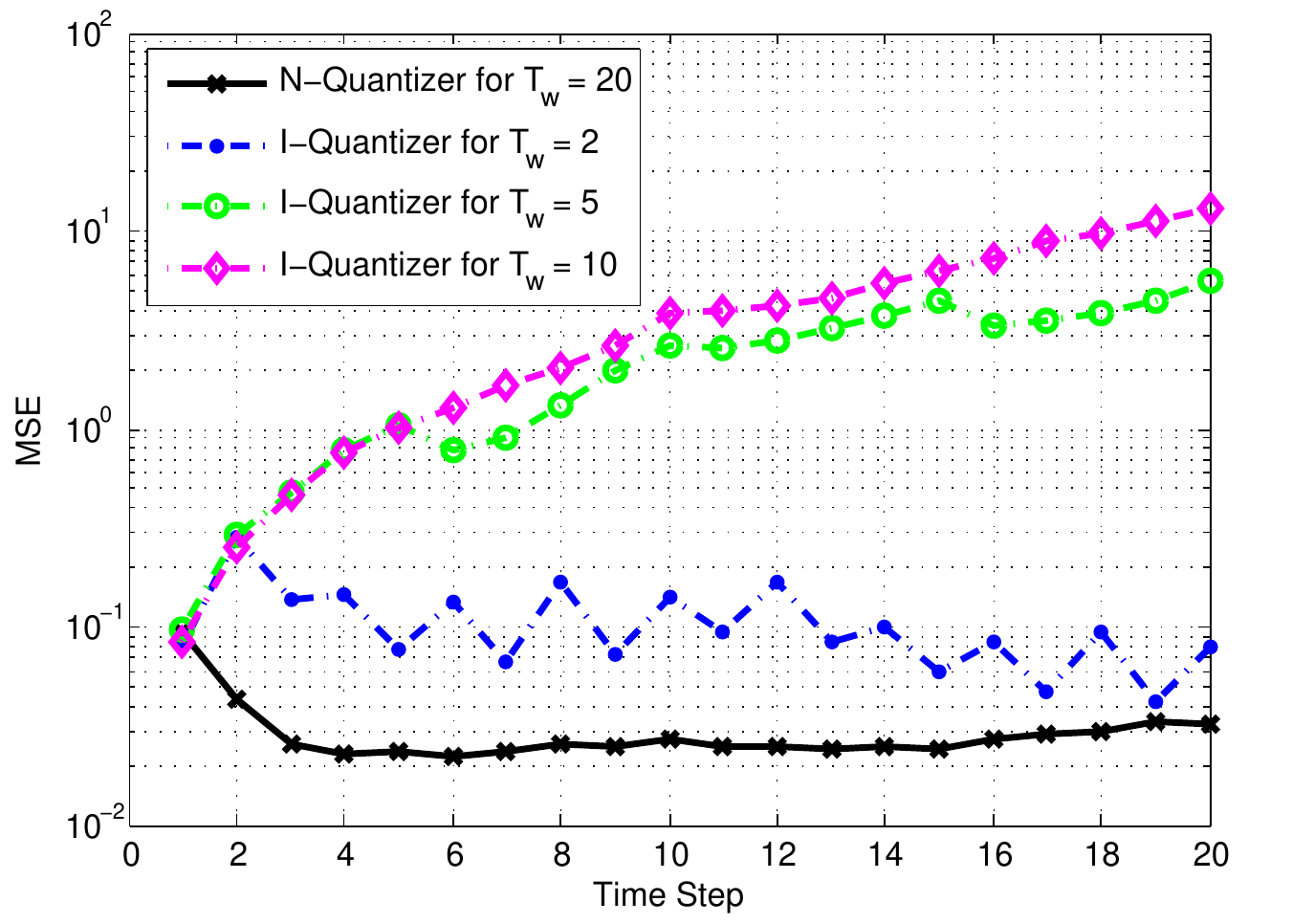}
\caption{\footnotesize{MSE performances of $2$-bits quantizer as $q = 2.5 \times 10^{-3}$}}
\label{fig: Iquantizer}
\end{figure}

In Fig.\,\ref{fig: Iquantizer}, we present the MSE of temporally identical quantizer (I-Quantizer), which refers to the design of quantizers only for the next time step and then using the same quantizers over the entire time window. For comparison, the MSE of nonidentical quantizer (N-Quantizer) with $T_w = 20$ is also plotted. Simulation results show that the
I-Quantizer yields worse performance than the N-Quantizer even with a small window size (i.e., $T_w =2$). This is because in a tracking scenario the target state is random and dynamic, which leads to a large innovation error by using temporally-identical quantizers, although the identical design can save energy and computation cost.


\section{Conclusion}
\label{sec: conclude}

In this paper, we considered the problem of target tracking with quantized data in a WSN, where the optimal local quantizers are determined using a non-myopic strategy. Using the alternative conditional posterior Cram\'er-Rao lower bound (A-CPCRLB) as the performance metric, we theoretically showed that the non-myopic quantizer can be designed separately for each time instant. With the help of a particle filtering method, this problem can be expressed in a closed form and solved via the interior-point algorithm. Simulation results demonstrated the effectiveness of our proposed approach. 
In the future, we will consider the effect of channel statistics on quantizer design. We will also consider a unified non-myopic optimization framework for resource management problems such as sensor selection and bit allocation.

\appendix

\textit{Proof of Proposition\,$1$}:
According to (\ref{eq: A_CBIM}), problem (\ref{eq: non_mop_CFIM}) can be decomposed into two subproblems
\begin{equation*}
\resizebox{.85\hsize}{!}{$
\displaystyle
\begin{array}{cl}
\displaystyle \maximize_{\bm \gamma_{t+T_w}} & E_{p_{t+T_w}^c} \left \{  - \nabla_{\mathbf x_{t+T_w}}^{\mathbf x_{t+T_w}}   \mathrm{ln} p (\mathbf u_{t+T_w}| \mathbf{x}_{t+T_w})\right \} \\
\text{subject to} &  \gamma_{t+T_w,1}^i < \cdots < \gamma_{t+T_w,L-1}^i, ~ \text{$i = 1, \cdots, N$}
\end{array}
$}
\end{equation*}
and 
\begin{equation}
\resizebox{.78\hsize}{!}
{$
\displaystyle
\begin{array}{cl}
\displaystyle \maximize_{\{\bm \gamma_{t+\eta}\}} & \left( Q + F J_{t+T_w -1}^{-1}F^T \right)^{-1} \\ 
\text{subject to} & \gamma_{t+\eta,1}^i < \cdots < \gamma_{t+\eta,L-1}^i \\
& \text{$\eta=1,\cdots, T_w-1$ and $i = 1, \cdots, N$}
\end{array}
$}
\label{eq: J_Tw_1}
\end{equation}
where $Q$ and $F$ are given by the process model (\ref{eq: state0}). 

Note that $ Q + F J_{t+T_w -1}^{-1}F^T $ is positive definite since $Q$ is positive definite and the information matrix $J_{t+T_w -1}$ is positive definite (where we assume it is invertible). Then,
problem (\ref{eq: J_Tw_1}) can be written as
\begin{equation}
\resizebox{.78\hsize}{!}{$
\displaystyle
\begin{array}{cl}
\displaystyle \minimize_{\{\bm \gamma_{t+\eta}\}} & Q + F J_{t+T_w -1}^{-1}F^T  \\ 
\text{subject to} & \gamma_{t+\eta,1}^i < \cdots < \gamma_{t+\eta,L-1}^i\\
& \text{$\eta=1,\cdots, T_w-1$ and $i = 1, \cdots, N$}
\end{array}
$}
\label{eq: J_Tw_2}
\end{equation}
where we use the fact that, for any positive definite matrix, if $A \succeq B$ then $B^{-1} \succeq A^{-1}$.

Since $F$ is invertible, 
the problem (\ref{eq: J_Tw_2}) is equivalent to
\begin{equation*}
\resizebox{.79\hsize}{!}
{$
\displaystyle
\begin{array}{cl}
\displaystyle \maximize_{\{\bm \gamma_{t+\eta}\}} & J_{t+T_w -1} \\ 
\text{subject to} & \gamma_{t+\eta,1}^i < \cdots < \gamma_{t+\eta,L-1}^i \\
& \text{$\eta=1,\cdots, T_w-1$ and $i = 1, \cdots, N$}
\end{array}.
$}
\end{equation*}

Similarly, after $T_w$ recursive decompositions, the problem (\ref{eq: non_mop_CFIM}) can be decomposed into $T_w$ sub-problems given by 
\begin{equation}
\resizebox{.8\hsize}{!}
{$
\displaystyle
\begin{array}{cl}
\displaystyle \maximize_{\bm \gamma_{t+\eta}} &   E_{p_{t+\eta}^c} \left \{  - \nabla_{\mathbf x_{t+\eta}}^{\mathbf x_{t+\eta}}   \mathrm{ln} p (\mathbf u_{t+\eta}| \mathbf{x}_{t+\eta})\right \} \\
\text{subject to} &  \gamma_{t+\eta,1}^i < \cdots < \gamma_{t+\eta,L-1}^i, ~ \text{$i = 1, \cdots, N$}
\end{array}
$}
\label{eq: decompose_nmop_ntr}
\end{equation}

Then by \cite[Lemma 3.1]{SP_2013}, the problem (\ref{eq: decompose_nmop_ntr}) is equivalent to
the problem (\ref{eq: decompose_nmop}), where for clarity , we reiterate the \cite[Lemma 3.1]{SP_2013} as below. 

Consider two optimization problems 

\begin{subequations}
\begin{align}
    \resizebox{.08\hsize}{!}{ $   \displaystyle \max_{\mathbf x \in \mathcal S} $}& \quad  \resizebox{.1\hsize}{!}{ $ M(\mathbf x) $}\tag{$A_1$} \label{eq: A1}\\
   \resizebox{.08\hsize}{!}{   $  \displaystyle \max_{\mathbf x \in \mathcal S} $} & \quad  \resizebox{.145\hsize}{!}{$ \mathrm{tr}( M(\mathbf x)) $} \tag{$A_2$} \label{eq: A2}
\end{align}
\end{subequations}
where $ M(\mathbf x)$ is a matrix for an arbitrary $\mathbf x \in \mathcal S$, $\mathcal S$ specifies the constraint on $\mathbf x$. If the problem (\ref{eq: A1}) has an optimal solution, then the problem (\ref{eq: A1}) is equivalent to (\ref{eq: A2}).

The proof of \cite[Lemma 3.1]{SP_2013} includes two parts. 
First it can be shown that if $\mathbf x_1$ is the
optimal solution of (\ref{eq: A1}), then for arbitrary $\mathbf x \in \mathcal S$, $M(\mathbf x) \preceq M(\mathbf x_1)$ which yields $\mathrm{tr}(M(\mathbf x)) \leq \mathrm{tr}(M(\mathbf x_1))$. Thus, $\mathbf x_1$ is also the optimal solution of (\ref{eq: A2}).
On the other hand, if $\mathbf x_2$ is the optimal solution of (\ref{eq: A2}), then we have $\mathrm{tr}(M(\mathbf x_1)) \leq \mathrm{tr}(M(\mathbf x_2))$. Note $\mathbf x_1$ is the optimal solution of (\ref{eq: A1}) which implies $\mathrm{tr}(M(\mathbf x_2)) \leq \mathrm{tr}(M(\mathbf x_1))$. Thus, we obtain $\mathrm{tr}(M(\mathbf x_1) - M(\mathbf x_2)) = 0$. By $\mathrm{tr}(M(\mathbf x_1) - M(\mathbf x_2)) = 0$ and $M(\mathbf x_1) - M(\mathbf x_2) \succeq 0$, we have $M(\mathbf x_1) = M(\mathbf x_2)$. Therefore, $\mathbf x_2$ is also the optimal solution of (\ref{eq: A1}). 
\hfill$\blacksquare$

\section*{Acknowledgment}
\footnotesize {This work was supported by U.S. Air Force Office of Scientific Research
(AFOSR) under Grant FA9550-10-1-0263 and FA9550-10-1-0458.}

\bibliographystyle{IEEEbib}
\bibliography{journal}

\begin{thebibliography}{10}

\bibitem{MRV2012}
E.~Masazade, R.~Niu, and P.~K. Varshney,
\newblock ``Dynamic bit allocation for object tracking in wireless sensor
  networks,''
\newblock {\em IEEE Trans. Signal Process.}, vol. 60, no. 10, pp. 5048--5063,
  Oct. 2012.

\bibitem{MMP2007}
M.~Vemula, M.~F. Bugallo, and P.~M. Djuric,
\newblock ``Particle filtering-based target tracking in binary sensor networks
  using adaptive thresholds,''
\newblock in {\em Proc. IEEE Int. Workshop on Comp. Advances in Multi-Sensor
  Adaptive Processing}, Dec. 2007, pp. 17--20.

\bibitem{MOHC2011}
M.~Mansouri, O.~Ilham, H.~Snoussi, and C.~Richard,
\newblock ``Adaptive quantized target tracking in wireless sensor networks,''
\newblock {\em Wireless Networks}, vol. 17, no. 7, pp. 1625--1639, Oct. 2011.

\bibitem{ORV2008}
O.~Ozdemir, R.~Niu, and P.~K. Varshney,
\newblock ``Adaptive local quantizer design for tracking in a wireless sensor
  network,''
\newblock in {\em Proc. the 42nd Asilomar Conf. Signals, Systems and
  Computers}, Oct. 2008, pp. 1202--1206.

\bibitem{NV2006}
R.~Niu and P.~K. Varshney,
\newblock ``Target location estimation in sensor networks with quantized
  data,''
\newblock {\em IEEE Trans. Signal Process.}, vol. 54, no. 12, pp. 4519--4528,
  Dec. 2006.

\bibitem{YORP2012}
Y.~Zheng, O.~Ozdemir, R.~Niu, and P.~K. Varshney,
\newblock ``New conditional posterior {C}ram\'er-{R}ao lower bounds for
  nonlinear sequential {B}ayesian estimation,''
\newblock {\em IEEE Trans. Signal Process.}, vol. 60, no. 10, pp. 5549--5556,
  Oct. 2012.

\bibitem{Boyd2004_bk}
S.~Boyd and L.~Vandenberghe,
\newblock {\em Convex Optimization},
\newblock Cambridge University Press, Cambridge, 2004.

\bibitem{SP_2013}
X.~Shen and P.~K. Varshney,
\newblock ``Sensor selection based on generalized information gain for target
  tracking in large sensor networks,''
\newblock {\em \textnormal{Arxiv preprint} http://arxiv.org/abs/1302.1616},
  2013.

\bibitem{ADA2004}
A.~S. Chhetri, D.~Morrell, and A.~Papandreou-Suppappola,
\newblock ``Efficient search strategies for non-myopic sensor scheduling in
  target tracking,''
\newblock in {\em Proc. the 38th Asilomar Conf. Signals, Systems and
  Computers}, 2004, vol.~2, pp. 2106--2110.

\bibitem{MNV2012_ciss}
E.~Masazade, R.~Niu, and P.~K. Varshney,
\newblock ``An approximate dynamic programming based non-myopic sensor
  selection method for target tracking,''
\newblock in {\em Proc. the 46th Annual Conf. Information Sciences and
  Systems}, March 2012, pp. 1--6.

\bibitem{LRV2011}
L.~Zuo, R.~Niu, and P.~K. Varshney,
\newblock ``Conditional posterior {C}ram\'er-{R}ao lower bounds for nonlinear
  sequential {B}ayesian estimation,''
\newblock {\em IEEE Trans. Signal Process.}, vol. 59, no. 1, pp. 1--14, Jan.
  2011.

\bibitem{MSN2002}
M.~S. Arulampalam, S.~Maskell, N.~Gordon, and T.~Clapp,
\newblock ``A tutorial on particle filters for online nonlinear/non-{G}aussian
  {B}ayesian tracking,''
\newblock {\em IEEE Trans. Signal Process.}, vol. 50, no. 2, pp. 174 --188,
  Feb. 2002.

\bibitem{HK_bk}
H.~L. Van~Trees and K.~L. Bell,
\newblock {\em Bayesian Bounds for Parameter Estimation and Nonlinear Filtering
  Tracking},
\newblock Wiley-IEEE press, 2007.

\bibitem{ABP2013}
A.~Vempaty, B.~Chen, and P.~K. Varshney,
\newblock ``Optimal quantizers for {B}ayesian distributed estimation,''
\newblock {\em Proc. IEEE Int. Conf. Acoustics, Speech and Signal Processing},
  2013.

\end{thebibliography}

\end{document}